\documentclass[twocolumn,floatfix,prl,showpacs,superscriptaddress]{revtex4-1}

\usepackage{graphicx,color,bm}
\usepackage{amsmath}
\usepackage[breaklinks,colorlinks,bookmarks=false,citecolor=blue,linkcolor=red,urlcolor=blue]{hyperref}

\newcommand{\abs}[1]{\left|#1\right|}

\usepackage{epsfig}

\begin{document}

\title{ESR modes in a Strong-Leg Ladder in the Tomonaga-Luttinger Liquid Phase }

\author{M.~Ozerov}
\thanks{Present Address: FELIX Laboratory, Radboud University, 6525 ED Nijmegen, The Netherlands}
\affiliation{Dresden High Magnetic Field Laboratory (HLD-EMFL), Helmholtz-Zentrum Dresden-Rossendorf, D-01328 Dresden, Germany}

\author{M. Maksymenko}
\affiliation{Max-Planck-Institut f\"{u}r Physik komplexer Systeme, D-01187 Dresden, Germany}
\affiliation{Institute for Condensed Matter Physics, National Academy of Sciences of Ukraine, L'viv-79011, Ukraine}
\affiliation{Department of Condensed Matter Physics, Weizmann Institute of Science, Rehovot 76100, Israel}

\author{J.~Wosnitza}
\affiliation{Dresden High Magnetic Field Laboratory (HLD-EMFL), Helmholtz-Zentrum Dresden-Rossendorf, D-01328 Dresden, Germany}
\affiliation{Institut f\"ur Festk\"orperphysik, TU Dresden, D-01062 Dresden, Germany}

\author{A.  Honecker}
\affiliation{Institut f\"ur Theoretische Physik, Georg-August-Universit\"at G\"ottingen, D-37077 G\"ottingen, Germany}
\affiliation{Laboratoire de Physique Th\'eorique et
Mod\'elisation, CNRS UMR 8089,
Universit\'e de Cergy-Pontoise, 
F-95302 Cergy-Pontoise Cedex, France}

\author{C.~P.~Landee}
\affiliation{Dept. of Physics and Carlson School of Chemistry, Clark University, Worcester, MA 01060, USA}
\author{M.~M. Turnbull}
\affiliation{Carlson School of Chemistry and Biochemistry, Clark University, Worcester, MA 01610, USA}

\author{S. C. Furuya}
\author{T. Giamarchi}
\affiliation{Dept. of Quantum Matter Physics, University of Geneva, CH-1211 Geneva, Switzerland}

\author{S.~A. Zvyagin}
\affiliation{Dresden High Magnetic Field Laboratory (HLD-EMFL), Helmholtz-Zentrum Dresden-Rossendorf, D-01328 Dresden, Germany}


\begin{abstract}

Magnetic excitations in the
strong-leg quantum spin ladder compound (C$_7$H$_{10}$N)$_2$CuBr$_4$ (known as DIMPY)
in the field-induced Tomonaga-Luttinger spin liquid phase are studied by means
of high-field electron spin resonance (ESR) spectroscopy. The presence of
a  gapped ESR mode with unusual non-linear frequency-field dependence is
revealed experimentally. Using a combination of analytic and exact
diagonalization methods, we compute the dynamical structure factor and
identify this mode with  longitudinal excitations in the antisymmetric
channel. We argue that these  excitations constitute a fingerprint of the spin dynamics
in a strong-leg spin-1/2 Heisenberg antiferromagnetic
 ladder   and owe their ESR observability to the uniform Dzyaloshinskii-Moriya
interaction.

\end{abstract}
\maketitle

The investigation of spin systems where quantum effects play a dominant role has become a very active branch of
quantum many-body physics.
Although the spin Hamiltonian describing quantum magnets is quite simple and
often very well controlled \cite{auerbach_book_spins}, the interplay of all spin   degrees of freedom can be very complex, leading  to a large diversity  of phases ranging from long-range magnetic order to spin liquids of various types \cite{balents_spin_liquids}. In addition,  the ground state can  possess not only local types of order but also more complex and subtle non-local topological orders \cite{haldane_gap,nijs_equivalence}. Understanding such behavior is thus a frontier of fundamental knowledge, providing, on the other hand,  a potential means for quantum computation \cite{nayak_tqc_review} or quantum simulators of some itinerant problems \cite{giamarchi_BEC_dimers_review,ward_simulator_review}.

Due to enhanced quantum effects,  one- and quasi-one-dimensional (1D) spin systems, such as spin chains and ladders,  are of particular interest   \cite{dagotto_ladder_review}. In these systems, interactions between  excitations can play a very important role,  giving  rise to  exotic states  \cite{giamarchi_book_1d}, including  quasi-long-range order, known as Tomonaga-Luttinger liquids (TLL),  or phases  where correlations between  magnetic excitations are of short range  (e.g., in the case of Haldane spin-1 chains  \cite{haldane_gap}).

Recent  progress in material science makes it possible  to synthesize new
materials with exchange parameters permitting   the manipulation of   the ground
states  by accessible magnetic fields, with drastic effects on the physical
properties. This, and the progress in both analytical and numerical techniques
provide  access to a host of novel physics, allowing, e.g., the observation
of the  Bose-Einstein condensation of magnons \cite{giamarchi_BEC_dimers_review,zapf_bec_review}, the quantitative test for TLL predictions \cite{klanjsek_nmr_ladder_luttinger,zheludev_timescaling}, the observation of fractionalization of spin excitations \cite{thielemann_spectrum_spinladder,bouillot_dynamics_ladder_DMRG_long}, spinon attraction \cite{schmidiger}, and remarkable effects of disorder \cite{hong_disorder_magnets,tanaka_disorder_tlcucl,zapfrocilde_disorder_bose_glass}.

Even very tiny  anisotropies can play an important role, reducing local symmetries and  drastically affecting the low-energy spin dynamics.
Electron spin resonance (ESR) spectroscopy has proven to be one of the most sensitive tools  to probe such interactions and effects  in exchange-coupled spin systems \cite{BG}.
One remarkable advantage of this technique is that ESR allows experiments in very high magnetic fields, far beyond the superconducting magnet limit
\cite{zvyagin_FEL,zvyagin_CuPM,ozerov_FEL_ESR,nojiri_CuGeO3}. Theoretical studies of predicted ESR parameters are available for spin chains and ladders \cite{oshikawa_epr_1,oshikawa_epr_2,oshikawa_epr_3,brockmann_esr,furuya_esr_imp,furuya_esr_width_ladder}, and have been applied with good success to, e.g., spin chains \cite{zvyagin_CuPM_1,zvyagin_CuPM_2, validov_zigzag_epr} and strong-rung ladders \cite{furuya_ESR_BPCB, cizmar_esr_bpcb}. However, relatively little is known about the spin dynamics in strong-leg  ladder systems, which can  be very  different from that in spin chains and strong-rung ladders  in terms of the spinon interactions.

\begin{figure}[t]
\begin{center}
\includegraphics[width=0.45\textwidth]{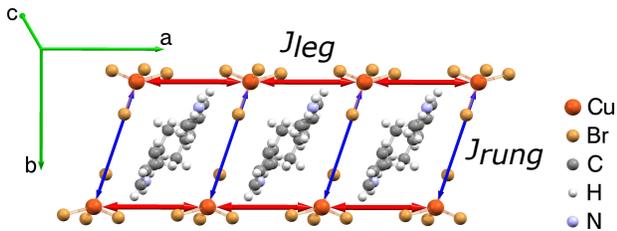}
\end{center}
\caption{\label{fig:structure} (color online)
Schematic view of the crystal structure of DIMPY~\cite{shapiro_chemistry}.
The copper (Cu) ions form a ladder-like structure with the dominant exchange couplings
indicated in the figure.
}
\end{figure}

In this work, we report on high-field ESR studies of the spin ladder (C$_7$H$_{10}$N)$_2$CuBr$_4$
[bis(2,3-dimethylpyridinium) tetrabromocuprate(II) or (2,3-dmpyH)$_2$-CuBr$_4$, abbreviated as DIMPY],
currently known as the best realization of a strong-leg spin-1/2 Heisenberg antiferromagnetic
 ladder \cite{shapiro_chemistry} with moderate
exchange coupling constants. We reveal experimentally the presence of a novel ESR excitation mode
in the TLL phase that is absent in a strong-rung ladder and was not observed in previous
ESR work on DIMPY either \cite{DIMPYesrGlazkov}.
We describe the unusual excitation spectrum of DIMPY, using a combination of analytic techniques
and exact-diagonalization (ED) methods. We demonstrate that the appearance and magnetic-field
dependence of parameters of the new mode can be understood by taking into account the dynamic spin-spin correlation
function for the strong-leg spin-1/2  Heisenberg antiferromagnetic ladder model, thus providing important information on the spin excitations as well as the anisotropy of magnetic interactions in this system.

DIMPY crystallizes in a monoclinic lattice with space group P2(1)/n and lattice constants $a=7.504 \AA$, $b=31.61 \AA$,  $c=8.202 \AA$, $\beta = 98.98^{\circ}$ (number of formula units per unit cell $Z=4$) ~\cite{shapiro_chemistry} with $S=1/2$ Cu$^{2+}$ ions
arranged in a ladder-like structure (Fig.~\ref{fig:structure}).
Each unit cell contains two rungs, each from a different symmetry-equivalent ladder, running parallel
to the $a$ axis.
The spin Hamiltonian of DIMPY can be written as
\begin{eqnarray}
\label{Ham}
\mathcal{H}  &=& J_{\rm leg}\sum_{\langle l,j \rangle} {\bf S}_{l,j}\cdot{\bf S}_{l+1,j}+J_{\rm rung} \sum_{\langle l \rangle}{\bf S}_{l,1}\cdot{\bf S}_{l,2}
\nonumber\\
&-&
g\mu_{B}H\sum_{l,j}S^{z}_{l,j}+\mathcal{H_{\delta}},
\end{eqnarray}
where $J_{\rm leg}$ and $J_{\rm rung}$ are exchange coupling constants along the legs and rungs, respectively,
${\bf S}_{l,j}$ are the spin operators on site $l$ of the leg $j=1,2$ of the ladder,
$g\mu_{B}H$ is the Zeeman term ($g$ is the $g$ factor, $\mu_B$ is the Bohr magneton, $H$ is the applied magnetic field). The
 fourth term represents various possible, usually small,
anisotropic contributions. Exchange constants along the rungs and legs of the ladder
have been determined by use of inelastic neutron scattering (INS) as $J_{\rm rung} /k_B \approx 9.5$~K
and $J_{\rm leg} /k_B \approx 16.5$~K, respectively ($J_{\rm leg}/J_{\rm rung}\sim 1.73$) \cite{schmidiger_dimpy_neutrons}.
The ladders are coupled via very weak exchange interactions, $J'/k_B \lesssim 5-7$~mK~\cite{shapiro_chemistry,schmidiger_dimpy_neutrons,DIMPYesrGlazkov}, resulting in a transition into a field-induced magnetically ordered phase at temperatures below $\sim 0.35$ K \cite{jeong_TTL_ladder}.

In a strong-leg ladder, the transverse interchain interaction couples two spin chains. As a result, two spinons are confined to magnons, opening a spin gap in the excitation spectrum. In the presence of a magnetic field the gap in DIMPY closes at a critical field $H_{c1}=2.8$~T, where the system undergoes a transition into the gapless TLL phase \cite{Hong}. Above $H_{c2}=29$~T, the system is in the magnetically saturated spin-polarized phase \cite{white_magnetization_dimpy}. Inelastic neutron scattering experiments revealed the presence of several gapless continua as well as a number of gapped excitations in DIMPY ~\cite{schmidiger_dimpy_neutrons,schmidiger_neutrons_magnetic,schmidiger_two_channels}; some of the excitations have been interpreted theoretically.
Investigating the field-induced evolution of the magnetic excitation spectrum  of a strong-leg ladder in  the TLL  state is  of particular interest,  so far not covered  in detail by theory and experiments. Such a study would allow to obtain a better understanding of peculiarities of  the spin dynamics in a strong-leg ladder in the TLL phase, which is, as shown below,  rather different from that known for quantum spin-1/2 chains and strong-leg ladders.

ESR experiments were performed at the Dresden High Magnetic Field Laboratory (Hochfeld Magnetlabor-Dresden), using transmission-type ESR spectrometers  (similar to that described in Ref.~\cite{spectrometer})  equipped with  $16$~T superconducting   and $50$~T pulsed-field \cite{Mike} magnets.  VDI modular transmitters (product of Virginia Diodes Inc., USA) and backward-wave oscillators (PO Istok, Russia) were employed as sub-mm radiation sources. High-quality single-crystal samples of DIMPY with typical sizes of $2\times1\times1~{\rm mm}^3$ were used in our experiments. The magnetic field was applied along the $b$ axis. In our experiments 2,2-diphenyl-1-picrylhydrazyl (DPPH) with $g=2.0036$ was used a standard ESR marker.

\begin{figure}[t]
\begin{center}
\includegraphics[width=0.5\textwidth]{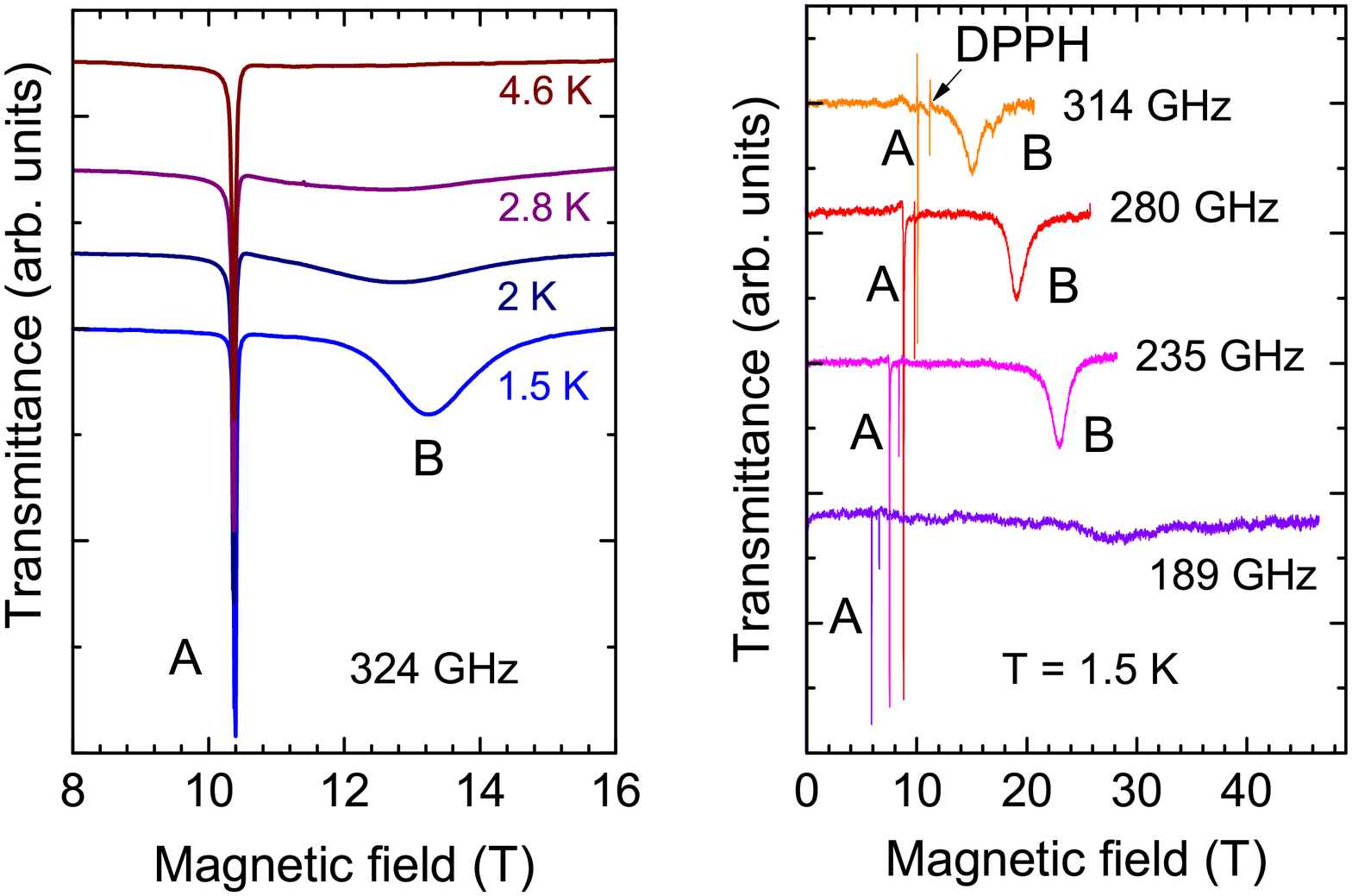}
\end{center}
\vspace{-0.5cm}
\caption{\label{Spectra} (color online) Left panel: Examples of ESR spectra obtained at a frequency of 324 GHz at 1.5, 2, 2.8, and 4.6 K.
Right panel:   Examples of pulsed-field ESR spectra  obtained at the frequencies 189, 235, 280, and 314.4 GHz ($T=1.5$ K).}
\end{figure}

\begin{figure}[t]
\begin{center}
\includegraphics[width=0.5\textwidth]{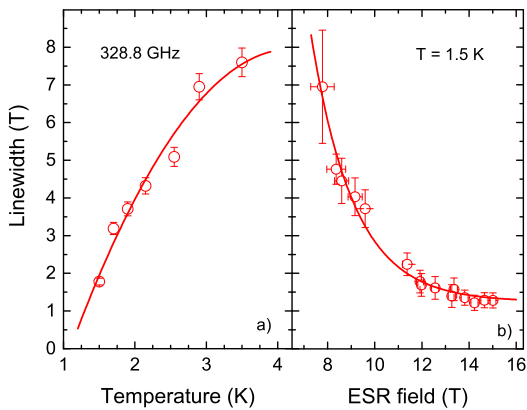}
\vspace{-0.8cm}
\end{center}
\caption{\label{Width} (color online) (a) Temperature dependence of the linewidth of Mode B at a frequency of 328.8 GHz. (b) Linewidth of Mode B for different values of resonance fields ($T=1.5$~K). Lines are guides to the eye. }
\end{figure}

A single resonance line (Mode A, Fig.~\ref{Spectra}) was observed at temperatures above $\sim 4$~K. At lower temperatures, the
ESR spectrum undergoes remarkable changes. In addition to Mode A we detected a relatively broad resonance absorption line (Mode B,
Fig.~\ref{Spectra}).   With decreasing temperature,  Mode B becomes more intensive and  narrower, shifting towards higher  fields.  Corresponding examples of the ESR spectra as well as the dependences of ESR linewidth (Mode B) on temperature and magnetic field are shown in Fig.~\ref{Spectra} and Fig.~\ref{Width}, respectively.

\begin{figure}[t]
\begin{center}
\includegraphics[width=0.5\textwidth]{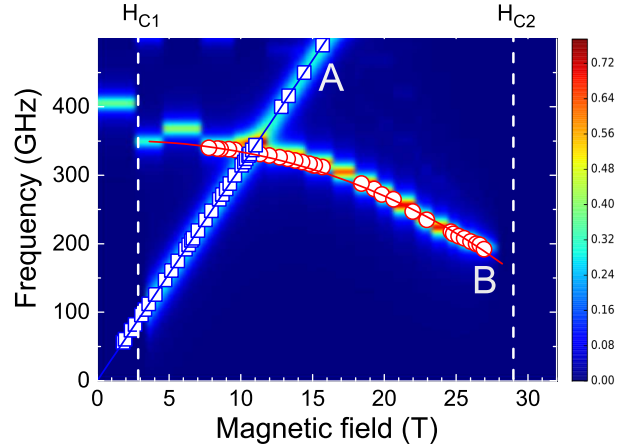}
\vspace{-0.8cm}
\end{center}
\caption{\label{FFD} (color online) The frequency-field diagram of magnetic
excitations in DIMPY.
Data of the structure factors $S^{zz}_{\pi}$ and $S^{\pm}_{0}$ obtained by use
of exact-diagonalization calculations  for chains from $N=32$ to $N=64$ sites
are given in bright colors~\cite{Suppl}. Blue and red solid lines are guides to the eye. First and second critical fields are denoted by vertical dashed  lines.  }
\end{figure}

The frequency-field diagram of the magnetic excitations in DIMPY is shown in
Fig.~\ref{FFD}.  Mode A (white boxes in Fig.~\ref{FFD}) can be described using
the equation $h \nu =g_b\mu_B H$, where $h$ is the Planck's constant, $\nu$ is
the excitation frequency, and $g_b=2.23$.  Mode B (white circles in
Fig.~\ref{FFD}) has a more complex behavior: this mode is gapped for all fields and has a non-linear frequency-field dependence. From  the extrapolation of the frequency-field dependence to zero field,  the energy gap, $\Delta\sim 350$~GHz,  can be  estimated. This value agrees   well with the size of the gap between the spin-singlet ground and first-excited triplet states  observed by means of INS in zero magnetic field  at $k=0$ \cite{schmidiger_neutrons_magnetic,schmidiger_two_channels}, where the system is in the gapped spin-liquid state.


It is worth mentioning that the ESR excitation spectrum  in DIMPY is very different from that in the strong-rung spin ladder BPCB \cite{cizmar_esr_bpcb}, where only one gapless mode was observed in the TLL phase.  The comparison of our ESR data with results of INS studies and ED calculations for DIMPY in the TLL regime \cite{schmidiger_neutrons_magnetic} strongly suggests  that the observed ESR Mode A corresponds to magnetic excitations in the $S^{\pm}_{0}$  channel, while  Mode B corresponds to  ESR excitations in the channel $S^{zz}_{\pi}$, which are nominally forbidden in the purely isotropic case.
To demonstrate that $S^{zz}_{\pi}$ indeed gives rise to Mode B,
we calculated the field dependence of the
dynamical structure factor employing ED of the model ({\ref{Ham}}),
where the anisotropic contribution $\mathcal{H_{\delta}}$ has been omitted.
We used the parameters $J_{\rm leg}/J_{\rm rung} = 1.73$, $J_{\rm rung} = 9.51$~K,
and $g=2.23$ as determined above from the frequency-field dependence of Mode A \cite{fn1}.
The transverse dynamical structure factor $S^{\pm}_{0}$
in the symmetric channel of the legs and
the longitudinal dynamical structure factor $S^{zz}_{\pi}$
in the antisymmetric channel
are calculated for finite systems of up to $64$ sites using the expression
\begin{eqnarray}
S^{\alpha\beta}_{k_\perp}(\omega)= \frac{1}{\pi}
\sum_{n} \operatorname{Im}\frac{
\langle 0|S^\alpha_{k_\perp}|n\rangle\,
\langle n|S^\beta_{k_\perp}|0\rangle}
{\omega-(\epsilon_{n}-\epsilon_{0}+i\,\eta)}\,,
\label{eq:Ssum}
\end{eqnarray}
for $T=0$, where $|n \rangle$ are the eigenstates with energy $\epsilon_{n}$
($|0 \rangle$ is the ground state). $\eta$ is a Lorentzian broadening that we
set to $\eta=0.05\,J_{\rm rung}$.
The Fourier-transformed spin operators are given by
\begin{equation}
{S}^\alpha_{k_\perp} = \frac{1}{\sqrt{N}}\sum_{l,j}
\exp(i\,{k_\perp}\,j)\,{S}^\alpha_{l,j}\,.
\end{equation}
Thus, ${k_\perp}$ is the momentum perpendicular to the ladder, while we have assumed zero momentum along the ladder direction,
as is common for ESR.

We exploit the conservation of total $S^z$ of the model ({\ref{Ham}}).
When the dimension of the subspace is sufficiently small, we use full
diagonalization to evaluate (\ref{eq:Ssum}) while for bigger dimensions
we use first the Lanczos algorithm \cite{Lanczos,DagottoRevModPhys66} to find the ground state $|0 \rangle$ and
then a continued-fraction expansion \cite{DagottoRevModPhys66,HHK72,GBPhysRevLett59} to
obtain the spectral function (\ref{eq:Ssum}).

Our ED results for the zero-temperature dynamical structure factors $S^{\pm}_{0}$ and $S^{zz}_{\pi}$
are shown as intensity plots in Fig.~\ref{FFD}.
Finite-size effects are strongest for low magnetic fields where they may amount to errors of up
to 50~GHz for Mode B~\cite{Suppl}. For magnetic fields $H > 12.5$~T, the main finite-size effects
are the steps observed in ``line'' B and thus one may estimate them to not exceed 10~GHz
here. The agreement of the position of the intensity maxima for $H\gtrsim7$~T with the experimental ESR and INS \cite{schmidiger_neutrons_magnetic} data is excellent, including not only the downward slope,
but also the curvature of Mode B.
We note further that application of ED to finite temperature \cite{Suppl} also reproduces
the qualitative trends observed in Fig.~\ref{Spectra} and Fig.~\ref{Width} (a), in particular
a substantial thermal broadening of Mode B.

As mentioned, ESR transitions corresponding to  Mode B are nominally forbidden in the purely isotropic case.
On the other hand, Fig.~\ref{fig:structure} shows that there is
no inversion center on bonds along the ladder legs in the crystal
structure of DIMPY and successive tetrabromocuprate units are related by unit cell translations \cite{shapiro_chemistry}.
This allows for the presence of a uniform Dzyaloshinskii-Moriya (DM) interaction
along the legs of the form $\sum_l \sum_{j=1,2}
(-1)^j{\bf{D}}\cdot{(\bf{S}}_{l,j}\times{\bf{S}}_{l+1,j})$. It is important to mention that the uniform DM interaction has been found responsible
for a number of unusual effects, including, e.g., broadening of resonance line A, observed in DIMPY by means of low-frequency ESR spectroscopy \cite{DIMPYesrGlazkov} and the zero-field gap opening in the triangular-lattice antiferromagnet Cs$_2$CuCl$_4$ \cite{Starykh}. On the other hand, such a term accounts for the intensity of Mode B,
that is directly proportional to the spin-spin correlations as discussed above  \cite{Suppl}.
In the low-field limit, Mode B can be described using the non-Abelian bosonization approach~\cite{Shelton_ladder}, where it is understood as a complex of two Majorana fermions~\cite{Suppl}.
The magnetic field couples symmetrically to the two legs of the ladder whereas
the Majorana fermions are antisymmetric under the exchange $j=1\leftrightarrow j=2$ of the two legs.
Thus, to first approximation, Mode B is not affected by the applied field.
This accounts for the almost flat behavior of Mode B  observed in Fig.~\ref{FFD}
up to about 15~T \cite{Suppl}. At higher magnetic fields, renormalization effects of these Majorana fermions
are more important,  resulting in the observed non-linear frequency-field dependence of Mode B.

Our observations of the $S^{zz}_{\pi}$ mode can have broader impact in the context of the $SO$(5) ladder model \cite{Scalapino_SO5, furusaki_SO5}. In this model, the quantum phase transition driven by the chemical potential can  be mapped to the field-induced phase transitions in the Heisenberg ladder. In that case, the gapless excitations in the TLL state of spin ladders (Mode A) are interpreted as massless $t_{i+}$ bosons, while the gapped excitations (Mode B) correspond to massive $t_{i,0}$ bosons~\cite{furusaki_SO5}. The former contribution is characteristic of the TLL state and is commonly found in spin-1/2 Heisenberg chains (and can be interpreted as originating from the Bose condensate of $\Delta S^z=1$ magnons), while the latter have $\Delta S^z=0$  magnons  as their origin.  The boson mass is determined  by the Luttinger constant $K$ (describing the nature of interactions between particles) and the velocity  $u$; both parameters are field dependent \cite{schmidiger_dimpy_neutrons}. The complex contributions of these two variables  to the gapped excitation give a hint for understanding the non-linear dependence of Mode B in a magnetic field  as observed in our experiments.

To summarize, the excitation spectrum in DIMPY, a spin-1/2 Heisenberg antiferromagnetic strong-leg ladder compound, was probed by means of high-field ESR in magnetic fields up to 50 T. Two ESR modes were observed.
One of them  has a linear frequency-field dependence, and corresponds to Zeeman-split massless
$S_{0}^{\pm}$ excitations, commonly found in spin-1/2 Heisenberg chains and strong-rung ladders in the Tomonaga-Luttinger liquid regime.
On the other hand, we show that a key property of the ESR spectrum in a spin-1/2 Heisenberg strong-leg ladder  in the TLL phase
is the presence of gapped $S^{zz}_{\pi}$ excitations that derive from the gapped $\Delta S^z=0$
boson.
Good agreement between results of exact-diagonalization calculations and the experimental data was demonstrated.

\begin{acknowledgments}

This work was partially supported by the Helmholtz Gemeinschaft via the Virtual Institute ``New states of matter and their excitations'', Deutsche Forschungsgemeinschaft (DFG, Germany), Swiss SNF under Division II, and ERC synergy UQUAM project.  We acknowledge the support of the HLD at HZDR, member of the European Magnetic Field Laboratory (EMFL).

\end{acknowledgments}


\begin{center}
\Large{\textbf{Supplemental Material}}
\end{center}

\section{Uniform Dzyaloshinskii-Moriya interaction}

\label{sec:Bos}

Here, we show that a uniform Dzyaloshinskii-Moriya (DM) interaction along chains but staggered between chains
can explain that the dynamical structure factor
$S^{zz}_\pi$ leads to an ESR signal and that there is,
consequently, a nontrivial Mode B.

The symmetry of the crystal structure of DIMPY (Fig.~1 of the main manuscript) allows the occurrence of a uniform DM
interaction along the leg, but staggered from leg to leg
\begin{equation}
 \mathcal{H}_{\rm{DM}}=\sum_l\sum_{j=1,2}(-1)^j{\bf{D}}\cdot{(\bf{S}}_{l,j}\times{\bf{S}}_{l+1,j}),
  \label{HDM}
\end{equation}
that corresponds to the term $\mathcal{H_{\delta}}$ of the Hamiltonian (1) in the main text.
We choose coordinates in spin space such
that the vector ${\bf{D}}$ points in the $y$ direction,
${\bf{D}}=D\hat{y}$, where $\hat{y}$ is the unit vector along the $y$
axis.

Since the DM interaction is very small, it hardly affects most physical
quantities. One noteworthy exception is the selection rule of ESR excitations.
The mode allowed in the ESR spectrum strongly depends on weak
perturbations breaking the spin-rotational symmetry.
Normally, the ESR experiment measures $S^\pm_0$ at $k=0$.
The dynamical structure factor
\begin{equation}
 S^\pm_0(\omega)=-\frac
  1{1-e^{-\omega/T}}\operatorname{Im}G^R_{S^+S^-}(\omega)
  \label{Spm2Gpm}
\end{equation}
is proportional to the imaginary part of the retarded Green's function
at $k=0$,
\begin{equation}
 G^R_{S^+S^-}(\omega)
  = -i\int_0^\infty dt  e^{i\omega t} \langle [S^+(t), S^-(0)]\rangle,
  \label{Gpm}
\end{equation}
where $S^+=\sum_l\sum_{j=1,2}S^+_{l,j}$.
Since the prefactor in Eq.~\eqref{Spm2Gpm} does not have special resonances in frequency,
the ESR modes  A and  B mentioned in  the main manuscript come from the frequency dependence of the retarded Green's function
\eqref{Gpm}.
In order to investigate the origin of these modes, we utilize the following \emph{identity}
(compare with the appendix of Ref.~\cite{oshikawa_epr_4})
\begin{align}
 G^R_{S^+S^-}(\omega)
 &= \frac{2\langle S^z \rangle}{\omega-g\mu_BH} -
 \frac{\langle[\mathcal{A}, S^-]\rangle}{(\omega-g\mu_BH)^2}
 \notag \\
 & \quad
 +\frac
 1{(\omega-g\mu_BH)^2}G^R_{\mathcal{A}\mathcal{A}^\dagger}(\omega) \, ,
 \label{id}
\end{align}
where $\mathcal{A}$ is
the operator $\mathcal{A}=[\mathcal{H}_{\rm{DM}}, S^+]$.
According to the identity \eqref{id}, in the absence of the DM
interaction (i.e., $\mathcal{A}=0$), one finds a single mode at
$\omega=g\mu_BH$, which is Mode A.
The ESR mode B is absent unless anisotropic interactions
breaking the spin-rotational symmetry are present.
The identity \eqref{id} also shows that Mode  B comes from the
last term, the retarded Green's function
$G^R_{\mathcal{A}\mathcal{A}^\dagger}(\omega)$.
The operator $\mathcal{A}$ is given by
\begin{equation}
 \mathcal{A}=\sum_l\sum_{j=1,2}(-1)^jD
  (S^y_{l,j}S^x_{l+1,j}-S^x_{l,j}S^y_{l+1,j}).
  \label{A}
\end{equation}
The above formula is exact and valid for arbitrary magnetic field.

We can get an approximation valid for low energy, long wavelength
by replacing $S^{x,y}_{l+1,j} \to S^{x,y}_{l,j}$. In that case
Eq.~\eqref{A} becomes $\mathcal{A}=-iD\sum_l\sum_{j=1,2}(-1)^jS^z_{l,j}$.
and one obtains
\begin{equation} \label{eq:mapzz}
S^{zz}_\pi\approx -D^{-2}\,(1-e^{-\omega/T})^{-1}
\operatorname{Im}G^R_{\mathcal{A}\mathcal{A}^\dagger}(\omega)
\end{equation}
Mode  B is thus directly connected to the $zz$ spin spin correlation
function in the absence of DM interactions.

In order to complement the above description of Mode B we employ a bosonization approach dealing with the rung interaction,
$J_{\rm rung}$, perturbatively.
The bosonization provides a perfectly controllable treatment of the low-energy theory for $J_{\rm rung}/J_{\rm leg}\ll 1$.
Although for DIMPY the ratio is too high to expect quantitative agreement we can expect a very good qualitative description.

The standard non-Abelian bosonization calculation leads to~\cite{Gangadharaiah}
\begin{equation}
 \mathcal A \approx \frac{D}{\pi\alpha} \int dx
  \sum_{j=1,2}(-1)^j[J^z_{jR}(x)- J^z_{jL}(x)],
  \label{A_J}
\end{equation}
where $\alpha$ is the short-distance cutoff.
$J_{jR}(x)$ and $J_{jL}(x)$ are, respectively, the right-moving and the
left-moving components of the $SU(2)$ current on the $j$th leg.
Note that the spin ${\bf{S}}_{l,j}$ is written as
${\bf{S}}_{l,j}\approx
{\bf{J}}_{jR}(x)+{\bf{J}}_{jL}(x)+(-1)^{l+j}{\bf{N}}(x)$, where
${\bf{N}}(x)$ is the N\'eel order parameter.
According to Ref.~\cite{Shelton}, we can rewrite the operator
\eqref{A_J} by using two Majorana fermions,
\begin{equation}
 \mathcal A \approx \gamma D \int dx \, (\xi_R\rho_R-\xi_L\rho_L).
  \label{A_MF}
\end{equation}
Here, $\gamma$ is a non-universal constant.
$\xi_{R(L)}$ and $\rho_{R(L)}$ are the right-moving (left-moving)
component of the Majorana fermions $\xi$ and $\rho$.
The Majorana fermion $\xi$, when it is applied to the singlet ground
state at zero magnetic field, generates a triplon with $S^z=0$.
Although Ref.~\cite{Shelton} formulated the refermionized theory for the $H=0$ case, one can easily extend it to the high-field case of our interest. At the level of the bosonized and refermionized theory, the Hamiltonian of the spin ladder at $H=0$ is split into two parts: a symmetric and an antisymmetric part with respect to the permutation of the leg index $j=1\leftrightarrow j=2$. The magnetic field affects the symmetric sector only and can induce
a quantum phase transition from the gapped phase at low field into the field-induced Tomonaga-Luttinger liquid phase.
Conversely the Majorana fermions $\xi$ and $\rho$ belong to the antisymmetric sector~\cite{Shelton}, are thus in first approximation
unaffected by the magnetic field, and retain a gapful excitation spectrum. If we call $\Delta_0$ the excitation gap of $\xi$ at $k=\pi$, the other Majorana fermion $\rho$ has a higher excitation gap
$3\Delta_0$ at $k=0$~\cite{Shelton}. Hence, the operator \eqref{A_MF} generates multi-particle excitations whose excitation gap at $k=0$ equals to $4\Delta_0$.
For zero magnetic field the triplet gap $\Delta_0$ is estimated from our exact diagonalization
data as $\Delta_0\approx 4.3$~K (see section \ref{sec:EDT0} below) leading to a value
$4\Delta_0 \approx 17.2~{\rm K} \approx 360$~GHz
of the resonance frequency of Mode B
in good agreement with  $\Delta\sim 350$~GHz obtained from the extrapolation of the frequency-field dependence of the Mode B to zero field (see Fig.~4 in the main text). At finite magnetic field the
decoupling of the symmetric and antisymmetric mode would naively yield a field-independent frequency. However there are irrelevant operators that couple these two sectors.
Although they do not change the asymptotic physics they can renormalize the value of the parameters, hence a field dependence of the resonance that must be computed
numerically.

\section{Exact diagonalization of finite ladders}

\subsection{Zero temperature}

\label{sec:EDT0}

The field dependence of the dynamical structure factors in the $S^{\pm}_{0}$  and $S^{zz}_{\pi}$
channels at $k=0$ along the legs has been computed at $T=0$ using
exact diagonalization (ED) of the model Hamiltonian
[Eq.\ (1) in the main text] for finite systems of sizes up to $64$ sites.
For $N\le 20$ and in the high $S^z$ sectors we use full diagonalization,
otherwise a combination of the Lanczos algorithm with a continued fraction expansion
\cite{Lanczos,HHK72,GBPhysRevLett59,DagottoRevModPhys66}.

The action of the operators $S^{\pm}$ changes the quantum number $S^z$ by
$\pm 1$, rendering the computation of  $S^{\pm}_{0}$  more challenging
than $S^{zz}_{\pi}$.
Nevertheless, only a strong line at frequency $\omega = g\,\mu_B\,H$
has been observed in $S^{\pm}_{0}$ with an intensity subject to small finite-size effects.
Since furthermore the mixing of the two channels depends on parameters like
the length of the DM vector $D$ that are not really known, we manually
added a line for $S^{\pm}_{0}$  with a suitable intensity
in Fig.~4 of the main text.

\begin{figure}[tb!]
\begin{center}
\includegraphics[width=\columnwidth]{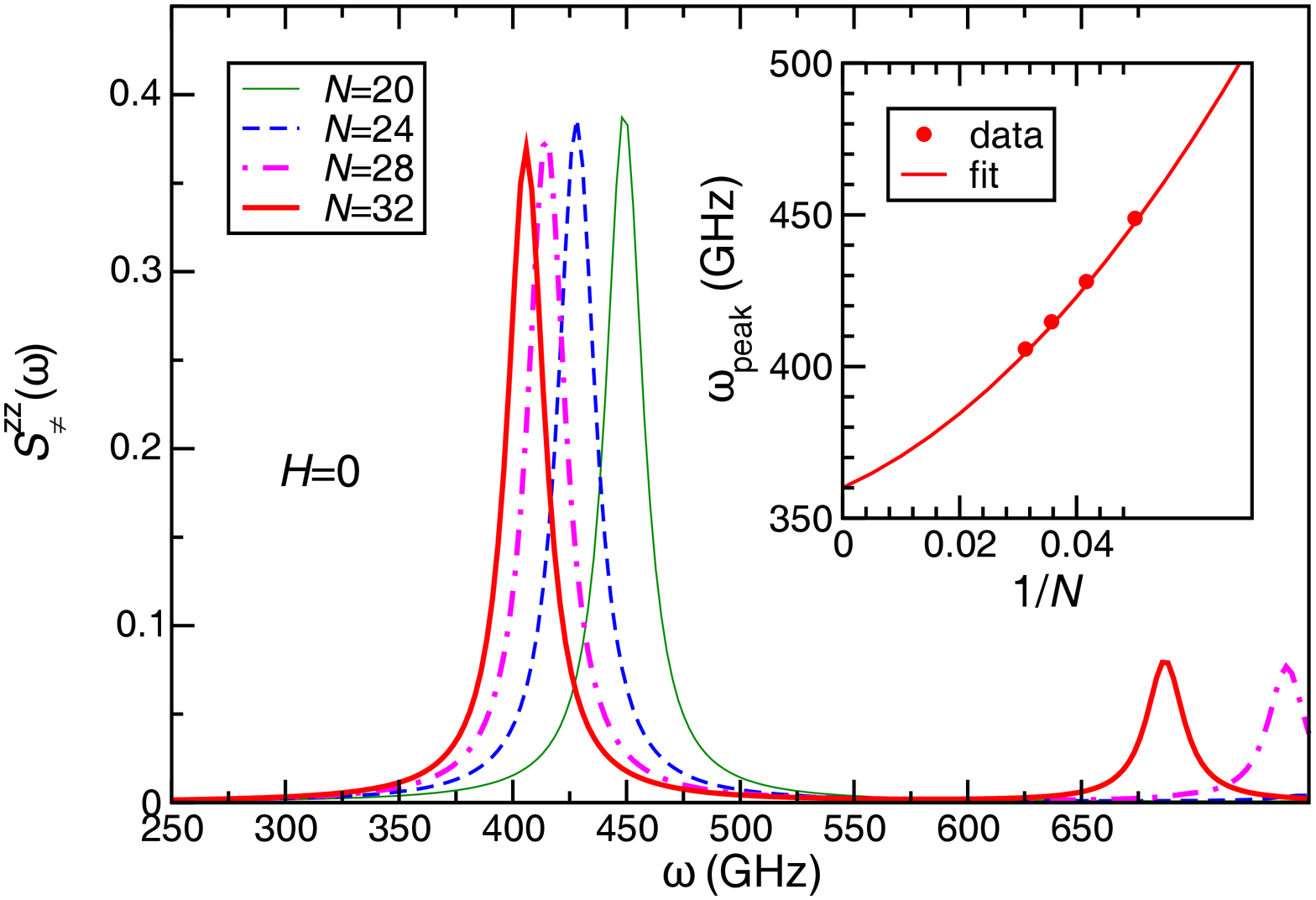}\\[1mm]
\includegraphics[width=\columnwidth]{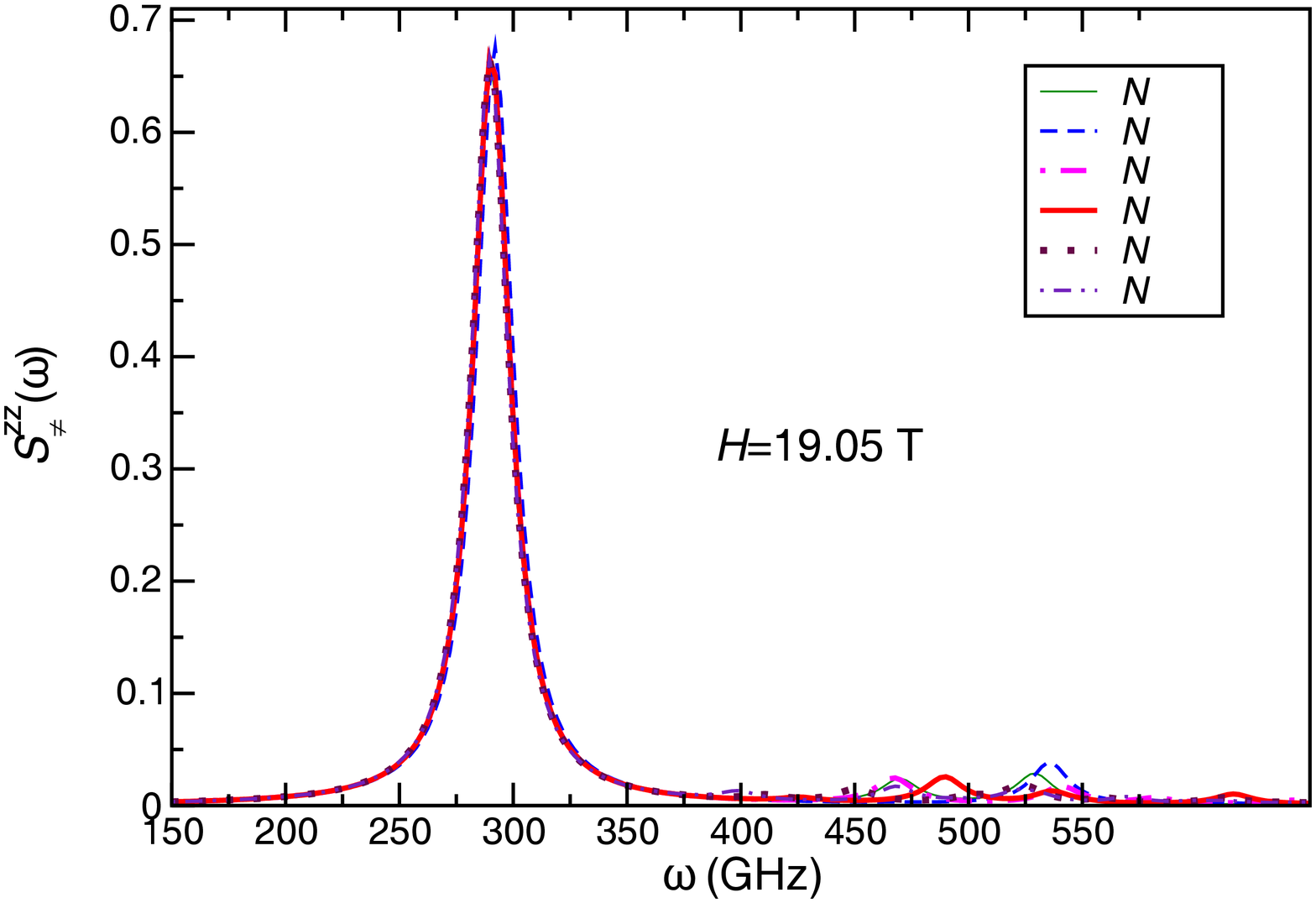}\\[1mm]
\includegraphics[width=\columnwidth]{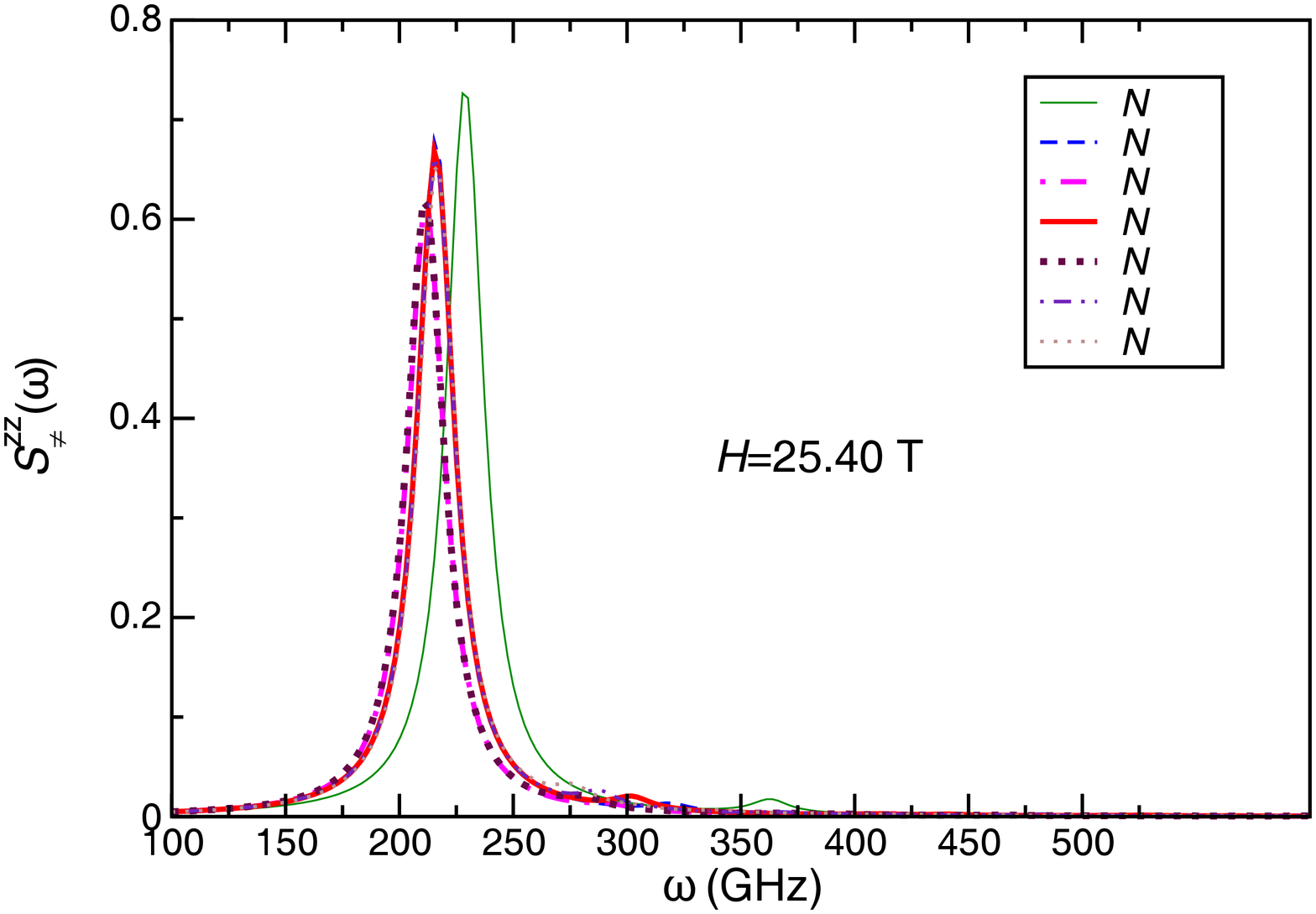}
\end{center}
\caption{(color online) Finite-size dependence of the $S^{zz}_{\pi}$ structure factor at the temperature $T=0$ and magnetic fields $H=0$
(top panel), $H=19.05$~T (middle panel), and $H=25.40$~T (bottom panel).
A Lorentzian broadening of $\eta = 0.05\,J_{\rm rung} \approx 10$~GHz is applied.
The inset of the top panel shows a finite-size extrapolation
of the frequency of the peak $\omega_{\rm peak}$ using a quadratic
fit in the inverse system size $1/N$.
\label{ED_size}
}
\end{figure}

We now focus on the channel $S^{zz}_{\pi}$ since this exhibits a more
complex behavior. For completeness, we start with the case of zero magnetic
field in the top panel of Fig.~\ref{ED_size} even if this has been investigated previously \cite{schmidiger,schmidiger_dimpy_neutrons,schmidiger_two_channels,schmidiger_neutrons_magnetic}.
The low-field region is particularly challenging for ED since on the one hand the numerical effort
is maximal and on the other hand finite-size effects are largest, compare the position of the
main peak in Fig.~\ref{ED_size}. Nevertheless, the inset of the top panel of Fig.~\ref{ED_size}
demonstrates that the peak position $\omega_{\rm peak}$ at $H=0$ can be
extrapolated to $\Delta \approx 360$~GHz in the thermodynamic limit $N\to \infty$.
Note that a similar extrapolation can also be performed for the spin gap $\Delta_0$: a fit with an exponential function
gives rise to $\Delta_0 \approx 4.3~{\rm K} \approx 90$~GHz, corresponding to a first critical field
$H_{c1} \approx 2.9$~T. As a consistency check of these two independent extrapolations, we mention that the ratio
$\Delta/\Delta_0$ reproduces the field-theory prediction $\Delta/\Delta_0 = 4$ very accurately.

\begin{figure}[htb!]
\begin{center}
\includegraphics[angle=270,width=\columnwidth]{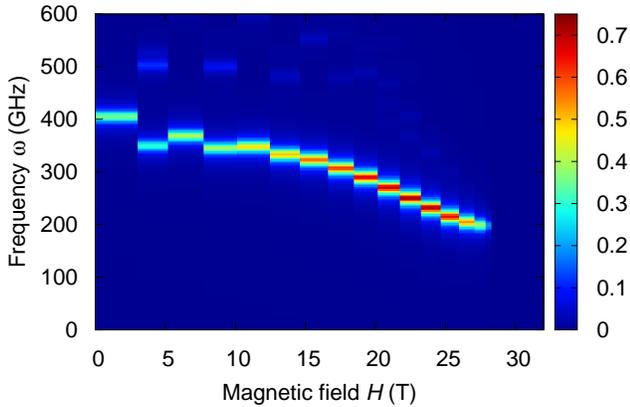}
\end{center}
\caption{(color online) Frequency-field diagram of
a finite ladder of $N=32$ sites.
The shading encodes the value of the structure factor $S^{zz}_{\pi}$.
\label{N32_appendix}
}
\end{figure}

The higher magnetic fields that are our main concern are more favorable for two reasons.
Firstly, exact diagonalization is performed for lower particle numbers and one can reach bigger system sizes.
In addition, finite-size effects become less important. The case $g\mu_BH=3J_{\rm rung}$ shown
in the middle panel of Fig.~\ref{ED_size} demonstrates a well-behaved case. In this case
the ground state is in the sector $S^z = N/4$ (half of the saturation magnetization)
for all considered systems and finite-size
effects are virtually absent, as is demonstrated by the lines for all system sizes
falling on top of each other in the middle panel of Fig.~\ref{ED_size}.

Finally, the bottom panel of Fig.~\ref{ED_size}
illustrates the more typical behavior with the case $g\mu_BH=4J_{\rm rung}$. In this
case, the ground state is in the sector $S^z = 3\,N/8$ for those $N$ that are divisible
by $8$, i.e., $N=24$, $32$, $40$, and $48$ in the figure. For these system sizes, again no finite-size
effects are observable. On the other hand, if $N$ is not divisible by $8$, $S^z = 3\,N/8$ cannot
be realized and the corresponding system sizes ($N=20$, $28$, and $36$ in the figure) scatter a bit around the thermodynamic
limit. For systems with $N\ge 32$ spins and in the high-field region, these finite-size shifts of
the main line should not exceed $10$~GHz.
Above this main peak there is always a bit of spectral density corresponding to continua of excitations.
Since these continua need to be approximated by a finite number of peaks for fixed $N$, one naturally
observes that these continua are more strongly affected by finite-size effects.

Figure~\ref{N32_appendix} shows the field dependence of $S^{zz}_{\pi}$  for a fixed system size of $N=32$ spins.
The jumps of the ``line'' in the low-field region in Fig.~\ref{N32_appendix} reflect again the
finite-size effects discussed above, but for higher magnetic fields the main effect is that
only discrete values of $S^z/N$ are realized for a fixed system size.
Fig.~4 of the main text is based on a composite of the largest available
system sizes and coincides with the present Fig.~\ref{N32_appendix} in the region $H<13.65$~T
(in Fig.~4 of the main text we have used $N=36$ for
$H>13.65$~T, $N=40$ for $H>18.86$~T, $N=48$ for $H>23.17$~T, and $N=64$ for $H > 26.03$~T).
Note that $S^{zz}_{\pi}(\omega) \equiv 0$ in the sector with $S^z =N/2 -1$ just before
saturation. In the case $N=32$ shown in the present Fig.~\ref{N32_appendix} this
implies a vanishing signal already at a field of $\approx 27.9$~T, i.e., below the saturation field $H_{c2} = 29$~T.
In the case of Fig.~4 of the main text we have used data for $N=64$ spins just below saturation.
Accordingly, this apparent saturation field is closer to the true saturation field, namely at $\approx 28.2$~T.

\subsection{Finite temperature}

\begin{figure}[tb!]
\begin{center}
\includegraphics[width=\columnwidth]{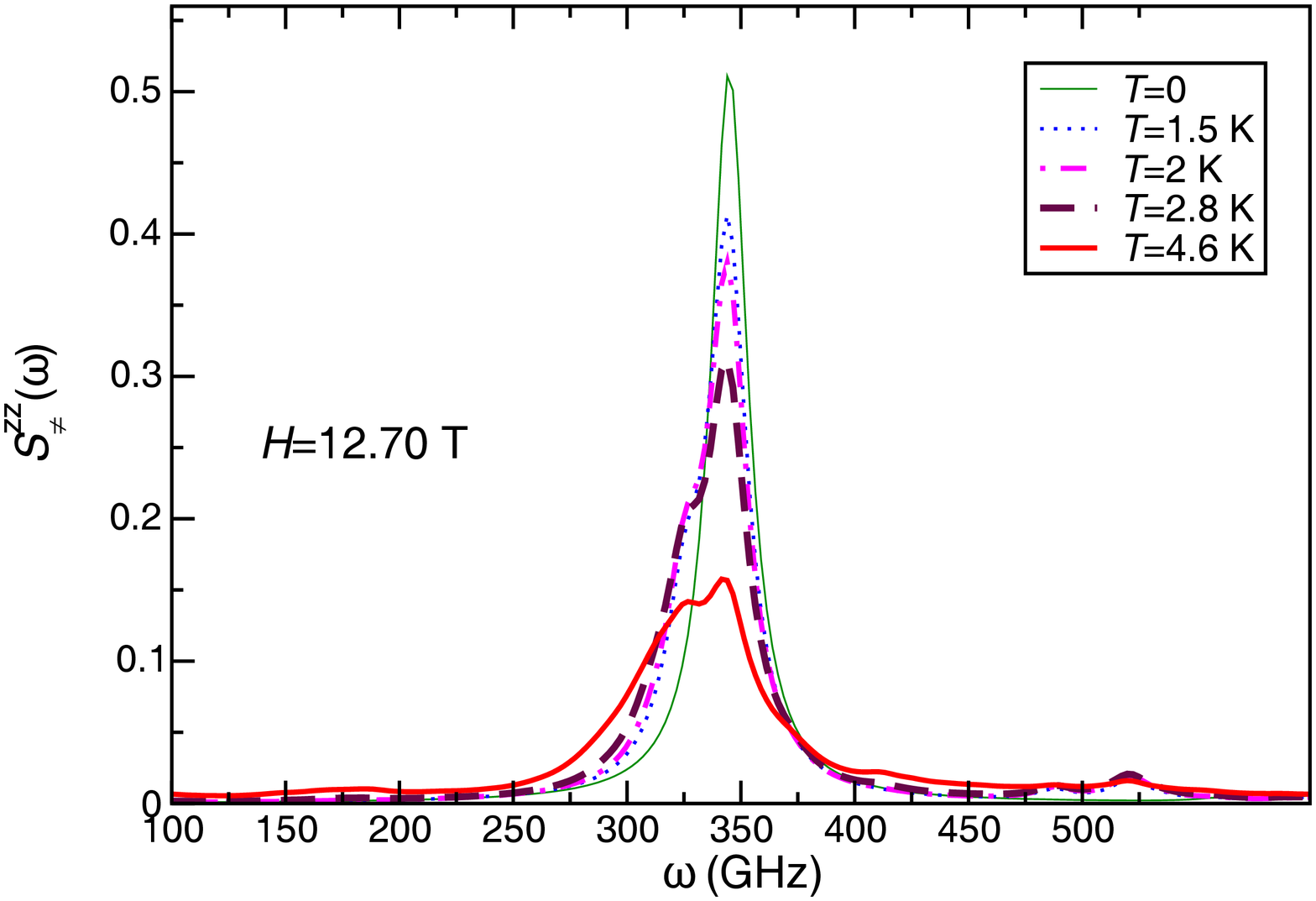}\\[1mm]
\includegraphics[width=\columnwidth]{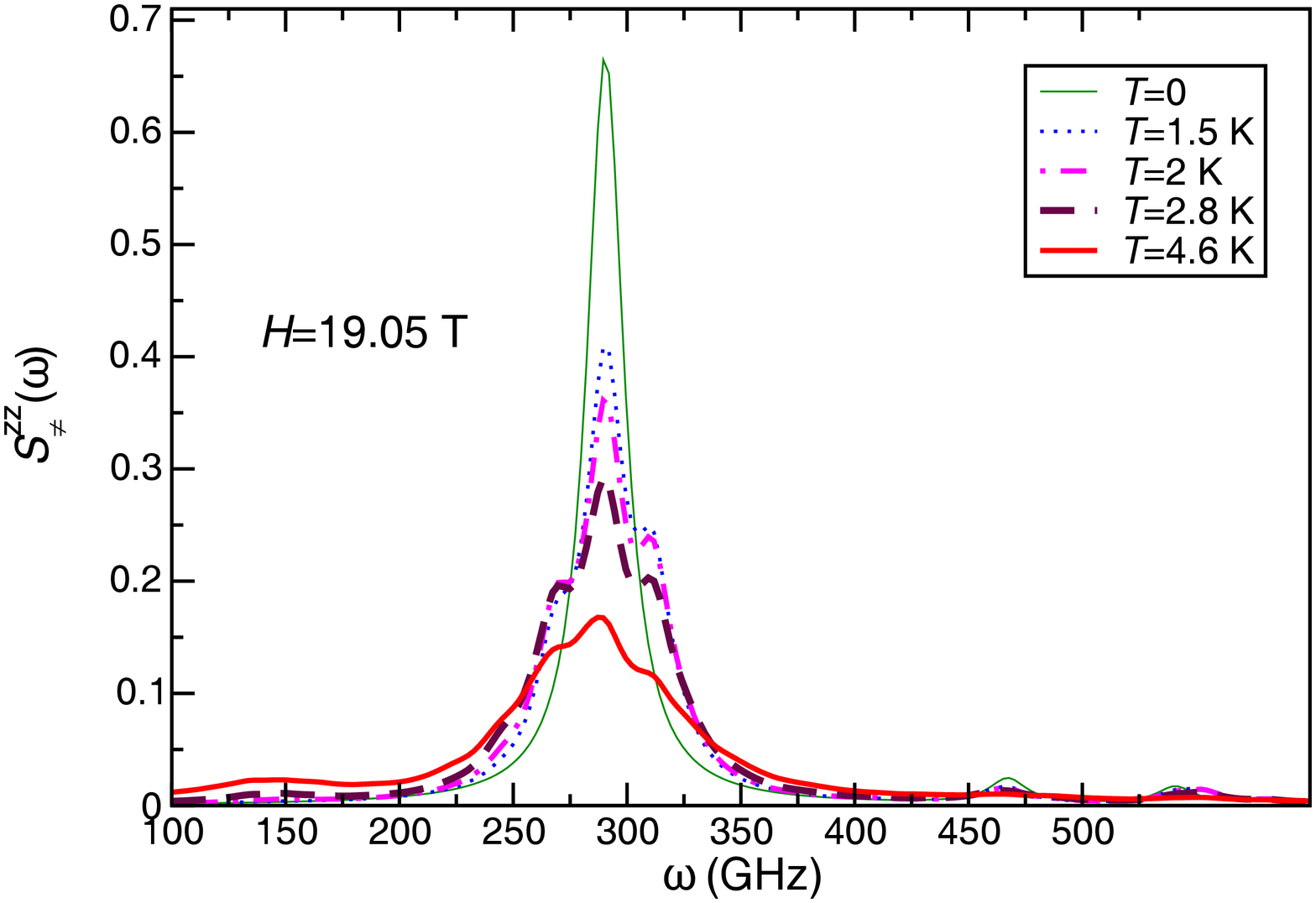}
\end{center}
\caption{(color online) Temperature dependence of the $S^{zz}_{\pi}$ structure factor at a magnetic fields
$H=12.70$~T (upper panel) and $H=19.05$~T (lower panel).
The size of the system is fixed at $N=28$ sites and a Lorentzian broadening
of $\eta = 0.05\,J_{\rm rung} \approx 10$~GHz is applied.
\label{fig:SzzPiT>0}
}
\end{figure}

Finally, we take a brief look at $T>0$. First, we need a generalization of Eq.~(2) of the main text
\begin{eqnarray}
S^{\alpha\alpha}_{k_\perp}(\omega)= \frac{1}{\pi}
\sum_{n,m}
\frac{e^{-\frac{\epsilon_m}{k_B\,T}}}{Z} \,
 \operatorname{Im} 
\frac{
\abs{\langle n|S^\alpha_{k_\perp}|m\rangle}^2}
{\omega-(\epsilon_{n}-\epsilon_{m}+i\,\eta)}\,,\quad
\label{eq:SsumTf}
\end{eqnarray}
where $Z = \sum_m e^{-\epsilon_m/(k_B\,T)}$ is the partition function.
Here we present ED results for $N=28$ sites. In this case, we can no longer obtain the full
spectrum, but we need to restrict the sums over $m$ to low energies $\epsilon_{m}$.
After performing such a restriction, the spectral sum over $n$ is again evaluated by a
continued fraction expansion \cite{DagottoRevModPhys66,HHK72,GBPhysRevLett59}.
This approximation would break down at high temperatures, but
is accurate for the region of interest, i.e.,
temperatures $k_B\,T < J_{\rm rung}/2$. Since we need to compute excited states
with the Lanczos algorithm \cite{Lanczos,DagottoRevModPhys66},
we have to work with smaller systems than for zero temperature.
At least 20 states have been retained for each sector with a given $S^z$ and momentum $\vec{k}$.

Figure \ref{fig:SzzPiT>0} presents results for two cases, namely $H=2\,J_{\rm rung}/(g\,\mu_B) \approx 12.70$~T
(upper panel) and $H=3\,J_{\rm rung}/(g\,\mu_B) \approx 19.05$~T (lower panel). The $T=0$ limit
of the latter case has been presented before in the middle panel of Fig.~\ref{ED_size}
where finite-size effects were observed to be small. At finite $T$, finite-size effects
are still visible as wiggles in the detailed lineshape. Still, Fig.~\ref{fig:SzzPiT>0}
clearly demonstrates a substantial broadening and corresponding damping as temperature
is raised to 4.6~K. This is in qualitative agreement with the experimental findings
(left panel of Fig.~2 and 3(a) of the main text). At $H=3\,J_{\rm rung}/(g\,\mu_B) \approx 19.05$~T there
is no observable shift of the position of the line with temperature whereas
for $H=2\,J_{\rm rung}/(g\,\mu_B) \approx 12.70$~T one observes a shift of the center of mass
of the line to lower frequency $\omega$ with rising temperature, at least for the $N=28$ system shown
in Fig.~\ref{fig:SzzPiT>0}. Given the downward slope of Mode B with increasing magnetic field $H$,
this translates to a shift of the mode to lower fields with rising temperature when one translates
the present frequency scans at constant field to field scans at constant frequency.
Thus, the shift observed in the upper panel of Fig.~\ref{fig:SzzPiT>0} is consistent
with the shift of Mode B observed experimentally (left panel of Fig.~2 of the main text).
Between $T=0$ and 1.5~K there is generally a bit of broadening, but no significant shift
of the main line, justifying the comparison of $T=0$ computations with experiments performed at
1.5~K.

\end{document}